# From Social Netizens to Data Citizens:

# Variations of GDPR Awareness in 28 European Countries

Răzvan Rughiniș, Department of Computer Science and Engineering, University POLITEHNICA of Bucharest, Bucharest, 060042, Romania, razvan.rughinis@cs.pub.ro

Cosima Rughiniș, Department of Sociology, University of Bucharest, Bucharest, 030167, Romania, cosima.rughinis@sas.unibuc.ro

[1]Simona Nicoleta Vulpe, Interdisciplinary School of Doctoral Studies, University of Bucharest, Bucharest, 050107, Romania, simona.vulpe@drd.unibuc.ro

Daniel Rosner, Department of Computer Science and Engineering, University POLITEHNICA of Bucharest, Bucharest, 060042, Romania, daniel.rosner@cs.pub.ro

## Abstract

We studied variability in General Data Protection Regulation (GDPR) awareness in relation to digital experience in the 28 European countries of EU27-UK, through secondary analysis of the Eurobarometer 91.2 survey conducted in March 2019 ($N$ = 27,524). Education, occupation, and age were the strongest sociodemographic predictors of GDPR awareness, with little influence of gender, subjective economic well-being, or locality size. Digital experience was significantly and positively correlated with GDPR awareness in a linear model, but this relationship proved to be more complex when we examined it through a typological analysis. Using an exploratory k-means cluster analysis we identified four clusters of digital citizenship, across both dimensions of digital experience and GDPR awareness: the *off-line citizens* (22%), the *social netizens* (32%), the *web citizens* (17%), and the *data citizens* (29%). The off-line citizens ranked lowest in internet use and GDPR awareness; the web citizens ranked at about average values, while the data citizens ranked highest in both digital experience and GDPR knowledge and use. The fourth identified cluster, the social netizens, had a discordant profile, with remarkably high social network use, below average online shopping experiences, and low GDPR awareness. Digitalization in human capital and general internet use is a strong country-level correlate of the national frequency of the data citizen type. Our results confirm previous studies of the low privacy awareness and skills associated with intense social media consumption, but we found that young generations are evenly divided between the rather carefree social netizens and the strongly invested data citizens. In order to achieve the full potential of the GDPR in changing surveillance practices while fostering consumer trust and responsible use of Big Data, policymakers should more effectively engage the digitally connected yet politically disconnected social netizens, while energizing the data citizens and the web citizens into proactive actions for defending the fundamental rights to private life and data protection.

---

[1] Corresponding author:
Simona - Nicoleta Vulpe, 36-46 Mihail Kogălniceanu, room 221 first floor, Sector 5, Bucharest, 050107, Romania. E-mail: simona.vulpe@drd.unibuc.ro.

# 1 Introduction

The accelerated advance of digital connectivity, from the internet and Web 2.0 to the Internet of Everything, powered by Big Data and artificial intelligence (AI), has led to a radical transformation of conduct in all areas of life, from business and politics to soul-searching, dating, and intimacy. In the Fourth Industrial Revolution, data is the new oil but also the new carbon, becoming constitutive of human personhood (Cheney-Lippold 2017). The good, the bad, and the ugly of data-fueled predictions have led to increasing concerns with privacy, particularly in digital arenas. The Charter of Fundamental Rights of the European Union (EU) specifies the respect for private and family life (Art. 7) and the protection of personal data (Art. 8) and fundamental freedoms. These rights have been further operationalized in the EU's General Data Protection Regulation (GDPR), enforceable since May 25, 2018. How are Europeans handling these new political developments in the fast-forward world of digital capitalism?

In this study, we looked deeper into a 2019 EU-level representative survey on people's digital experiences and awareness of the GDPR, Eurobarometer 91.2 which was conducted in March 2019 (European Commission 2019). We explored the variability in people's engagement in the digital arena and in their attention to privacy policy. We compared insights derived from linear and typological analyses, discussing implications for shaping a European model of data citizenship.

*A structural challenge in protecting people's rights to private life and protection of personal data* involves the economic gains that are derived from extracting and processing data, at the intersection of *surveillance capitalism* (Zuboff 2019) and *surveillance culture* (Lyon 2019). These gains may be illicit, derived through the big business of identity theft (Roberds and Schreft 2009), monetized through credit card or healthcare insurance frauds, for example. At the same time, personalized digital marketing has entered the core of legitimate present day economic organization, through the impact of giants such as Google, Amazon, and Facebook. Their business innovations have led to the emergence of surveillance capitalism, focused on the added value created through what Zuboff (2015, 2019) has termed "prediction products," which are obtained from the behavioral surplus of large volumes of data. As personal data become impersonal assets, to be sold, bought, transferred, and consolidated in ever larger datasets which hold the key to accurate customer predictions and corporate profits (Nissenbaum 2010). Thus, businesses all along the "data food chain" (Nissenbaum 2019) are incentivized to harvest increasingly richer information about users and to convert it into predictions and profits. A convergence of social media and app-generated traces, together with credit score systems, in the context of cultural legitimacy sources such as the quantified self and the rating culture, generate surveillance infrastructures (Park and Humphry 2019) that, in turn, make possible novel forms of data extraction and data-based prediction and regulation.

Along with the demands of surveillance capitalism and its associated cultural institutions, we have witnessed in recent months the *resurgence of public health surveillance* in the context of the COVID-19 pandemic. The EU "Joint European Roadmap towards Lifting COVID-19 Containment Measures" (April 2020) recommends that member states create "a framework for contact tracing and warning with the use of mobile apps, which respects data privacy" (European Council 2020, p.7), while tech giants Google and Apple have collaborated to produce a free Bluetooth based application programming interface (API) to monitor contacts with infected people by May of 2020 (Greenberg 2020). Digital monitoring of citizens at various levels of compulsion, anonymity, and security is a likely feature of the new coronavirus normality (Calvo et al. 2020; de Montjoye and Houssiau 2020).

The structural incentives for extracting and aggregating more and more data about people's whereabouts, actions, and thoughts are compounded by individual-level challenges of protecting the rights to private life and protection of data. The most notable challenge has been termed the "privacy paradox" (Kayes and Iamnitchi 2017; Gerber et al. 2018), referring to a persistent difference between many people's highly professed valorization of privacy vs. a low demonstrated actionable value. That is, people consistently seem to ignore or downplay privacy risks and to give away personal information for various online benefits, while declaring different preferences.

The privacy paradox of apparently hypocritical valorization of personal data is further exacerbated by regular users' *inability to make sense of sophisticated privacy agreements* (Acquisti et al. 2013, 2015). Trusting in individual, informed, free choice over data disclosure is theoretically risky if not downright impossible, given the layered complexity of corporate business practices regarding data, the chains through which data are transferred to various organizations with multiple interests and policies, and the obfuscation introduced into such agreements through corporate jargon (Nissenbaum 2010). In the big picture of the citizens' perception of privacy and data protection in Europe, there is a combination of high public concern and considerable confusion, though with variations of clarity and actions taken (Hallinan et al. 2012).

*The emergence of social media* and the institutionalization of sharing details of one's personal and intimate life with wide networks of followers, digital "friends," and unknown observers, has radically altered expectations of what about our lives is publish-worthy (Nissenbaum 2010), and has brought about a novel structure of rewards and risks for online disclosure (Kayes and Iamnitchi 2017). Involvement in social media requires curation of one's digital presence and work for the presentation of self (Goffman 1959) in front of many others. Still, attention for human-readable presentation of self has little to do with attention for machine-readable data flows that are gathered or extracted from one's online behavior. Therefore, social media users often invest considerable time and attention in controlling their digital profiles, walls, and channels, but these control skills are not transferable for controlling backstage personal data flows elicited through cookies, tracers and apps and platforms' various forms of data sharing. The so-called digital natives, young people who are immersed in the community life made possible by social networks, are often aware of privacy troubles but unwilling to pay for enhanced privacy (Engels 2019). Downloading and installing apps is also a behavior that normalizes taking privacy risks, as users are unlikely to engage with reading privacy agreements and worrying over privacy, given the high prevalence of apps in daily life (Braghin and Vecchio 2017).

Experienced digital navigators may also face the "paradox of control" (Brandimarte et al. 2013), when gaining improved understanding and abilities to configure a digital medium leads to higher self-efficacy, lower attention to privacy, and more actual disclosure than previously. There is a large information asymmetry between digital service providers and individual users in eliciting, configuring, and clarifying information flows; therefore, providers can offer an illusion of enhanced control while still nudging personal data sharing. The paradox of control explains why digitally skilled people are covering a large spectrum on privacy concerns, from a carefree end to a highly concerned and action-oriented end. Based on an EU-level survey, the Eurobarometer 74.3 (2010) distinguished two types of digital experts: the digital natives, including those in the 15–24 age group and students, who are intense users of social networking sites and are low on privacy concerns, and the "digital initiates," who are often educated professionals and managers and who are at the highest end of privacy concerns (TNS Opinion & Social 2010). As we discuss later in our Results section, while the age differentiation is not as stereotypically neat as one might assume, there is indeed a broad variation of privacy awareness among people who are experienced in digital arenas.

Despite massive challenges in controlling flows of personal information through legitimate and illegitimate pipelines, storage facilities, prediction engines, all is not lost. There is significant flexibility in people's awareness of privacy issues and skills in dealing with them, which are increasing to match evolving practices and are diverse at any given time.

States, corporate and nonprofit organizations, experts, and regular people acting as consumers and citizens are engaged in a *division of moral labor to protect fundamental rights to private life and personal data*, while still gaining from the economic and scientific promises of Big Data (Tene and Polonetsky 2011). Actors in a multitude of roles coordinate their actions in regulated societies, thus shaping a variety of *de facto* forms of data citizenship. The European model has foregrounded privacy as a fundamental right and as a pillar of the Digital Single Market, designing the GDPR as an instrument of consumer trust and citizenship in a digitally-mediated world. Still, the GDPR is only as powerful as its uptake by the various actors that participate in social action – from companies that are required to adjust their business practices, to individuals and organizations invited to take advantage of these levers of control over personal information flows, to national and European administrations that need to operationalize and enforce the new rules.

With the increasing relevance of personal data and Big Data for business and policy decisions, we have now witnessed the emergence of *data citizenship* as a novel form of digital citizenship. Data citizenship conceptualizes the variety of forms of inclusion and exclusion, of formal rights, and actual participation experiences in a state organization, driven by data processing. The concept is useful to point out that people may be *excluded* (Lerman 2013) or *discriminated* against (Eubanks 2018) through their data traces (O'Neil 2017; Eubanks 2018), because of inequalities in the quantity and quality of the data they generate and also because of novel forms of intersectional online–offline disadvantage (Park and Humphry 2019).

The concept of data citizenship also captures the varieties of *individual participation* (Yates et al. 2020) and *activism* (Lerman 2013) in the data-powered society. A consistent effort of conceptualizing and quantifying data citizenship has been taking place at the University of Liverpool through the project "Me and My Big Data 2020" (Yates et al. 2020), distinguishing three dimensions: *data thinking* through critical thinking and using data to make sense of the world, *data doing* through abilities to control one's own data flows and to competently deal with incoming flows, and *data participation* through involvement in others' data-related safety and well-being, shaping the data practices at the community and broader levels. The "Me and My Big Data 2020" report also highlighted the *dual relationship between digital engagement and data citizenship*. The authors started from a classification of digital media use, identifying six types for which they examined data practices: the extensive political users, the extensive users, the social and media users, the general users, the limited users, and the non-users. Converging with the Eurobarometer 74.3 report, the Liverpool University 2018 survey (Yates et al. 2020) on UK citizens determined the low levels of privacy concern among the social and media users, despite their relatively high digital experience: "Social and media users have almost as limited an awareness of the use of data by platforms as limited users. At the same time, they have the least concern about data sharing and the least critical position on the data sharing practices of platforms. Ironically, they also still do not trust content they find in any media – but they are more likely than other groups to trust content shared by friends. Given that this group (17% of users) consists mainly of young people, with lower educational attainment from lower income households, we are concerned that they will remain disadvantaged in their data literacy into later life" (p. 46).

We have also seen the emergence of a new concern over privacy, namely *algorithmic awareness* – given that data are processed through algorithms that raise different yet related issues for democratic citizenship and free consumer choice. A 2018 web survey of Norwegian citizens (Gran et al. 2020) found currently low levels of algorithmic awareness, thus emphasizing the need to include not just data but also algorithms in the public debate over privacy and data protection.

The GDPR represents a significant change in the European data citizenship infrastructure, but its uptake occurs at the intersection of various privacy cultures and ongoing efforts to reshape data flows. Privacy-related norms, expectations, and practices are different not only among social strata, but also among countries, given systematic divergence in their digitalization policies and political priorities. A recent study of data protection legislation and policies across the EU found that member states varied widely in their implementation of privacy and personal data policies and in the intensity of political debate, media attention, and public initiatives on data protection (Custers et al. 2018).

Research into the impact of GDPR is still incipient, and there have been mixed outcomes. A study of Facebook's tagging of consumers with sensitive ad labels revealed that the GDPR had little effect (Cuevas et al. 2019), and the use of cookies to track users appeared to be relatively immune to GDPR (Sanchez-Rola et al. 2019; Hu and Sastry 2019), indicating that users have acquired or exercised little additional power in this respect (Urban et al. 2019). Research on improvement in privacy policies documents progress (Urban et al. 2019) but also stagnation (Becher and Benoliel 2020). The future success of the GDPR depends on large-scale user engagement and organized implementation across the EU countries.

This article is organized as follows: we first present our indicators and methods for the secondary analysis of the Eurobarometer 91.2 data. We then proceed to discuss results of a linear modelling of GDPR awareness as a function of sociodemographic attributes and digital experience. We then conduct an exploratory cluster analysis and compare the insights from the typological analysis with those from linear modelling. The last section concludes the article.

## 2   Methods

We studied variability in GDPR awareness and data citizenship in the EU27-UK, through a secondary analysis on the Eurobarometer 91.2 conducted in March 2019 (European Commission 2019). We started with a linear model of variation of GDPR awareness as a function of digital experience and sociodemographic variables. We then conducted an exploratory typological analysis, classifying respondents on the two interrelated dimensions of GDPR, awareness and digital experience, and we then discussed the resulting sociodemographic stratification of cluster membership.

The results were statistically representative for EU27-UK after sample weighting. The multistage, random probability sample (European Commission and Kantar Public 2019) included 27,524 respondents 15 years of age and older. In both linear and typological analyses, we used pairwise deletion of missing cases.

For the typological analysis, we ran an exploratory k-means cluster analysis using the statistical software package IBM SPSS Statistics v. 22 (IBM, Armonk, NY, USA). We tested several cluster versions and reported the results from the most theoretically relevant classification that captured both differences in degree and differences in configuration, distinguishing four types of digital citizenship at the level of the EU27-UK population.

For both the linear and the typological analyses, variable values were recoded by grouping values to achieve a balanced distribution of answers across the sample, avoiding outliers, and by reordering numerical values in ascending scales to enhance cluster comparability and interpretation. The answer distributions for the selected indicators at the EU27-UK level and at country levels are presented in the Eurobarometer publication (European Commission 2019).

Digital experience was assessed in the Eurobarometer by collecting responses related to participants' frequency of internet usage, social network usage, and online purchases. Internet usage was measured using the following items: "You use the internet at home, your home," "You use the internet on your place of work," "You use the internet on your mobile device (laptop, smartphone, tablet, etc.)," and "You use the internet somewhere else (school, university, cyber-café, etc.). Respondents could choose from the following response options: "Every day or almost every day," "Two or three times a week," "About once a week," "Two or three times a month," "Less often," "Never," or "No internet access (SPONTANEOUS)." We computed a summative index for these items, ranging from "0 Never/No access" to "2 Everyday/Almost every day." The social network usage was elicited by asking the question "How often, if at all, do you use social networks?" The response options were: "Every day or almost every day," "Two or three times a week," "About once a week," "Two or three times a month," "Less often," or "Never." We recoded the response scale as ranging from "0 Never" to "3 Daily." Another question related to digital experience was "How often, if at all, do you purchase goods or services online?" The scale of response options was the same as the previous one. After collapsing categories of responses, we used a response scale with categories from "0 Never" to "3 Weekly." Engagement with privacy settings as part of digital experience was measured as: "Have you ever tried to change the privacy settings of your personal profile from the default settings on an online social network?," with a dichotomous response option of "Yes" or "No."

General awareness of GDPR regulations was measured in the Eurobarometer by the following question: "Have you heard of the General Data Protection Regulation (GDPR), which came into force in 2018?" The response scale comprised the following options: "0 No," "1 Yes but you don't know exactly what it is," and "2 Yes and you know what it is." Respondents were then asked about their awareness of the specific rights stipulated by the GDPR: "The General Data Protection Regulation guarantees a number of rights. Have you heard of each of the following rights?" The corresponding items were as follows: "The right to access your data," "The right to object to receiving direct marketing," "The right to correct your data if it is wrong," "The right to have your data deleted and to be forgotten," "The right to have a say when decisions are automated," and "The right to move your data from one provider to another." For each provision, response options included "0 No," "1 Yes but you have not exercised it" and "2 Yes and you have exercised it." The indicators for general and specific rights awareness were included as such in the typological analysis. For the multivariate linear model, we summed the indicator for specific right awareness into a formative indicator of GDPR awareness, ranking from 0 (no awareness) to 12 (has heard of all specific rights and exercised them all), and we used it as a dependent variable in the multiple regression analysis.

## 3  Results

In order to understand variability in GDPR awareness, we first conducted a multivariate linear regression analysis examining its relationship to sociodemographic categories, occupation, and digital experience. We then went on to examine the configurations of digital experience and GDPR awareness through a cluster analysis, identifying four types of data citizenship.

### 3.1  Linear analysis of GDPR awareness

As expected and documented in univariate analysis (European Commission 2019), GDPR awareness, measured through a summative indicator of awareness for each GDPR right, was stratified across the expected sociodemographic variations of gender, age, formal education, occupation, material situation, and residential community size. Education and occupation could not be included simultaneously in the model because they were strongly intercorrelated leading to problems of multicollinearity; thus, we included education in Models 1 and 2, and occupation in Models 3 and 4. Also, we included precise age rather than age categories in Models 2 and 4, in order to avoid multicollinearity with occupation types, due to the strong association between age categories and the "student" and "retired" occupational categories.

A multiple regression model that included only sociodemographic individual-level variables explained about 13% of GDPR awareness in the total EU27-UK population, and all predictors had statistically significant associations (see Model 1 in Table 1 and Model 3 in Table 2). On average and when all other conditions were equal in these models, men had a slightly higher GDPR awareness than women, younger generations had a higher awareness than the generation of people aged 55+ (the reference value in the model), people who were still studying or have graduated after age 16 had higher awareness than people with less or no formal education; also, people in larger communities and feeling more secure economically were also more aware of the GDPR than the others. The largest effects in terms of standardized coefficients were produced by higher education and by being in the age category of 25–39 years, while gender made the smallest statistical difference.

The relationship between sociodemographic conditions and GDPR awareness was mediated by respondents' *daily interests and experiences*, such as occupation and digital experience. For example, a model that included occupation rather than education (see Model 3 in Table 2) indicated that, on average, with all other variables being equal, age remained the strongest statistical predictor (awareness was negatively correlated with age), but specific occupational types, in particular being a manager and being a white-collar worker, had relatively strong positive statistical effects.

While managers and white-collar workers are more likely to have heard of the GDPR by virtue of their work, personal digital experience is more strongly related to an interest in privacy and in GDPR regulation. The *inclusion of digital experience in the multivariate regression models in addition to sociodemographics led to an explained variance of about 24%*. In Model 2 (Table1) and Model 4 (Table 2), we can see that internet use, online purchases and attempts to change privacy settings on social media were significant predictors of GDPR awareness, other things being equal. However, more intense use of social media had a weak negative predictive value on GDPR awareness.

It is also noteworthy that in the models that included digital experience as a predictor of GDPR awareness, *gender was no longer statistically significant, and age partially lost its predictive power*. In Model 2, the age bracket 15–25 years was no longer different from the reference category of 55+ years as, regards GDPR awareness, when controlling for digital experience and the other sociodemographics. In Model 4, the respondent's age was no longer statistically significant when controlling for occupation types. This indicated that the influence of age or generation on GDPR awareness was almost entirely mediated by digital experience.

*Table 1 - Multiple regression models of variation in GDPR awareness. Dependent variable: A formative index of awareness of GDPR rights (0-12). Data source: Eurobarometer 91.2.*

| Variables in the model | Values | Model 1 Socio-demographics | | Model 2 Socio-demographics and digital experience | |
|---|---|---|---|---|---|
| | | Beta | Sig. | Beta | Sig. |
| (Constant) | | | .000 | | .232 |
| Gender | 0 = Man, 1 = Woman | .024 | .000 | .010 | .139 |
| Respondent age category | Respondent is aged 15-24 | .139 | .000 | .021 | .057 |
| | Respondent is aged 25-39 | .173 | .000 | .036 | .000 |
| | Respondent is aged 40-54 | .143 | .000 | .044 | .000 |
| | Respondent is aged 55 or more (Reference category) | N/A | N/A | N/A | N/A |
| Respondent's age when graduating formal education | No formal education or less than 15 (Reference category) | N/A | N/A | N/A | N/A |
| | Graduated school at age 16-19 | .212 | .000 | .151 | .000 |
| | Graduated school at age 20+ | .330 | .000 | .196 | .000 |
| | Still studying | .146 | .000 | .093 | .000 |
| Size of community | 1=Rural area, 2=Towns and suburbs/small urban areas, 3=Cities and large urban areas | .042 | .000 | .016 | .018 |

| Variables in the model | Values | Model 1 Socio-demographics | | Model 2 Socio-demographics and digital experience | |
|---|---|---|---|---|---|
| Difficulties paying bills | 0=Almost never / Never, 1=From time to time, 2=Most of the time | -.075 | .000 | -.040 | .000 |
| Internet use | 0=Never/No access, 1=Often/Sometimes, 2=Everyday/almost everyday | Not included in the model | | .235 | .000 |
| How often, if at all, do you use social networks? | 0=Never, 1=Weekly or less often, 2=Daily | | | -.016 | .042 |
| How often, if at all, do you purchase goods or services online? | 0=Never, 1=Less often than monthly, 2=Monthly, 3=Weekly | | | .145 | .000 |
| Have you ever tried to change the privacy settings of your personal profile from the default settings on an online social network? | 0=No, 1=Yes | | | .199 | .000 |
| **Adjusted R Sq.** | | **13.3%** | | **24.2%** | |

*Table 2 - Multiple regression models of variation in GDPR awareness, including occupation. Dependent variable: A formative index of awareness of GDPR rights (0-12). Data source: Eurobarometer 91.2.*

| Variables in the model | Values | Model 3 Socio-demographics and occupation | | Model 4 Socio-demographics, occupation and digital experience | |
|---|---|---|---|---|---|
| | | Beta | Sig. | Beta | Sig. |
| (Constant) | | | 0.000 | | .006 |
| Gender | 0 = Man, 1 = Woman | .015 | .010 | .005 | .451 |
| Size of community | 1=Rural area, 2=Towns and suburbs/small urban areas, 3=Cities and large urban areas | .044 | .000 | .013 | .049 |
| Difficulties paying bills | 0=Almost never / Never, 1=From time to time, 2=Most of the time | -.081 | .000 | -.033 | .000 |
| Respondent age | (years) | -.226 | .000 | -.007 | .549 |
| Respondent occupation | Self-employed | .094 | .000 | .068 | .000 |
| | Manager | .186 | .000 | .132 | .000 |
| | Other white-collar occupations | .129 | .000 | .108 | .000 |
| | Manual worker | .036 | .000 | .046 | .000 |
| | House person | -.006 | .385 | .009 | .258 |
| | Unemployed | -.011 | .119 | -.008 | .325 |
| | Student | .015 | .121 | .048 | .000 |
| | Retired (Reference category) | N/A | N/A | | |
| Internet use | 0=Never/No access, 1=Often/Sometimes, 2=Everyday/almost everyday | Not included in the model | | .265 | .000 |
| How often, if at all, do you use social networks? | 0=Never, 1=Weekly or less often, 2=Daily | | | -.030 | .000 |
| How often, if at all, do you purchase goods or services online? | 0=Never, 1=Less often than monthly, 2=Monthly, 3=Weekly | | | .138 | .000 |

| Variables in the model | Values | Model 3 Socio-demographics and occupation | Model 4 Socio-demographics, occupation and digital experience |
|---|---|---|---|
| Have you ever tried to change the privacy settings of your personal profile from the default settings on an online social network? | 0=No, 1=Yes | .200 | .000 |
| **Adjusted R Sq.** | | **13.2%** | **24.3%** |

Linear analysis is useful in order to observe the big picture, the broad correlational patterns for GDPR awareness, searching for consistency. Still, the linear perspective does not capture the diversity of configurations, particularly those configurations that are inconsistent, grouping individuals ranking high-low or low-high on the correlated dimensions (Rughiniş and Rughiniş 2014). A typological analysis is best suited to understand the finer-grained processes that compose the big picture of stratification, capturing atypical types that illuminate the diversity of causal processes.

## 3.2 Typological analysis of data citizenship

In line with the conceptualization of Yates et al. (2020), we examined data citizenship at the intersection of privacy awareness and digital experience. We used both dimensions simultaneously and constructed combined types. Our approach was useful in understanding the co-variations of use and awareness and to determine their points of convergence and divergence.

While, by and large, a more frequent and diverse digital experience was expected to lead to higher privacy and GDPR awareness, as empirically confirmed through linear analysis in Models 1–4, this was not, however, consistently the case. We conducted an exploratory cluster analysis by grouping individuals in similar profiles and taking into account the two correlated dimensions of digital experience (including general internet use, social network use, online shopping, and changing privacy settings online) and GDPR awareness (including general awareness and awareness for each specific right). We found that a 4-cluster classification captured both the differentiation in degree and the differentiation in configuration across the two correlated dimensions (see Table 3).

*Table 3 - Cluster description. Values represent cluster averages. Data source: Eurobarometer 91.2.*

| | Cluster: | **Off-line citizens** | **Social netizens** | **Web citizens** | **Data citizens** | **Total** |
|---|---|---|---|---|---|---|
| | Description: | Low use, low GDPR awareness | High SN use, low GDPR awareness | Average use, above average GDPR awareness | High use, high GDPR awareness | EU27+UK sample |
| Internet use | 0=Never/No access, 1=Often/Sometimes, 2=Everyday/almost everyday | **0.63** | 1.96 | 1.50 | 1.96 | 1.59 |
| How often, if at all, do you use social networks? | 0=Never, 1=Weekly or less often, 2=Daily | **0.51** | **2.77** | 0.35 | 2.75 | 2.11 |
| How often, if at all, do you purchase goods | 0=Never, 1=Less often than monthly, | .39 | **1.17** | 1.13 | 1.92 | 1.33 |

| | Cluster: | Off-line citizens | Social netizens | Web citizens | Data citizens | Total |
|---|---|---|---|---|---|---|
| or services online? | 2=Monthly, 3=Weekly | | | | | |
| (For social network users) Have you ever tried to change the privacy settings of your personal profile from the default settings on an online social network? | 0=No, 1=Yes | 0.19 | **0.45** | **0.36** | 0.74 | 0.57 |
| Have you heard of the General Data Protection Regulation (GDPR), which came into force in 2018? | 0=No, 1=Yes but you don't know exactly what it is, 2 = Yes and you know what it is | 0.41 | 0.85 | 1.28 | 1.58 | 1.04 |
| Have you heard of the right to access your data? | 0=No, 1=Yes but you did not exercise it, 2=Yes and you exercised it | 0.13 | 0.62 | 1.09 | **1.44** | 0.83 |
| Have you heard of the right to object to receiving direct marketing? | 0=No, 1=Yes but you did not exercise it, 2=Yes and you exercised it | 0.08 | 0.54 | 1.15 | **1.52** | 0.82 |
| Have you heard of the right to correct your data if it is wrong? | 0=No, 1=Yes but you did not exercise it, 2=Yes and you exercised it | 0.08 | 0.55 | 1.03 | 1.40 | 0.77 |
| Have you heard of the right to have your data deleted and to be forgotten? | 0=No, 1=Yes but you did not exercise it, 2=Yes and you exercised it | 0.07 | 0.47 | 0.96 | 1.32 | 0.71 |
| Have you heard of the right to have a say when decisions are automated? | 0=No, 1=Yes but you did not exercise it, 2=Yes and you exercised it | 0.02 | **0.24** | 0.68 | 1.00 | 0.49 |
| Have you heard of the right to move your data from one provider to another? | 0=No, 1=Yes but you did not exercise it, 2=Yes and you exercised it | 0.05 | 0.36 | 0.85 | 1.26 | 0.63 |
| Weighted N | | 6,098 | 8,818 | 4,685 | 7,923 | 27,524 |

| Cluster: | Off-line citizens | Social netizens | Web citizens | Data citizens | Total |
|---|---|---|---|---|---|
| % within the EU27 + UK sample | 22% | 32% | 17% | 29% | 100% |

We identified *three clusters that captured difference in degree.* These were linearly aligned on the low-high continuum of experience and awareness. We also identified *a fourth cluster that captured difference in configuration: a discordant type of high internet use and low GDPR awareness* (Table 3). The first cluster, labeled off-line citizens, was characterized by an average low digital experience and an associated low privacy awareness. The next cluster on the low-high continuum was that of the so-called web citizens, with average internet use and online purchases but low social network use, and an above average GDPR awareness on all indicators. At the other end of the intensity continuum, data citizens had high levels on both dimensions and all indicators of use and awareness. The discordant cluster included social netizens, people who are very frequent users of social networks but below average online shoppers, while having low GDPR awareness. Therefore, they combine the discordant positions of both high and low digital experiences with low awareness.

In Table 4, we can see how clusters differed, on average, regarding the upper values of digital experience and GDPR awareness. Only about one-quarter of offline citizens used the internet daily, as opposed to almost everybody in the social netizen and data citizen clusters. While one-third of data citizens did online shopping weekly, this only applied to about one-tenth of social netizens and web citizens, and a tiny fraction of off-line citizens. Almost all data citizens and social netizens used social networks weekly or daily, in comparison with about 15% of web citizens and a tiny fraction of off-line citizens. Differences in GDPR awareness were equally strong, with more than two-thirds of data citizens clearly aware of GDPR, as opposed to 45% of web citizens, and about only 7% of off-line citizens. The social netizens had a relatively low awareness of GDPR, especially given their remarkably high online participation, with less than one-quarter being clearly aware of what it represents.

As discussed previously, *high participation in social networks is a documented liability for privacy awareness*, because of its cultivation of concern with human-readable presentation of self rather than machine-readable personal data flows, and also through the paradox of control. Our empirical results indicated that, indeed, there was a configuration of high social media consumption and low privacy awareness found in the social netizen type, which was present to a considerable degree in all European countries. In counter-distinction, engagement with online shopping seemed to cultivate a more instrumental relationship with the digital arena, becoming a component in both clusters with higher GDPR awareness, namely the web citizens and the data citizens.

We also observed that *algorithmic awareness*, measured through the indicator "Have you heard of the right to have a say when decisions are automated?," was at the lowest level among all specific right awareness indicators. This confirmed the incipient status of this issue on the public agenda (Gran et al. 2020), as well as the need to make it a priority by including it more systematically.

*Table 4 - Distribution of high digital use and high GDPR awareness across clusters. Data source: Eurobarometer 91.2.*

| Cluster | Off-line citizens (cluster %) | Social netizens (cluster %) | Web citizens (cluster %) | Data citizens (cluster %) | Total (cluster %) |
|---|---|---|---|---|---|
| Uses the internet (almost) every day | 22.3 | 95.6 | 66.1 | 96.9 | 74.7 |
| Does online purchases weekly | 1.5 | 10.5 | 11.6 | 32.1 | 17.1 |

| Cluster | Off-line citizens (cluster %) | Social netizens (cluster %) | Web citizens (cluster %) | Data citizens (cluster %) | Total (cluster %) |
|---|---|---|---|---|---|
| Uses social networks weekly or daily | 2.2 | 99.9 | 15.6 | 97.9 | 73.6 |
| Has tried to change privacy settings (for social network users) | 19.0 | 44.5 | 36.2 | 74.0 | 56.7 |
| Has heard of the GDPR and knows exactly what it is | 7.2 | 22.5 | 45.0 | 67.9 | 36.2 |

Despite expectations of consistency generated by the correlation of age with GDPR awareness, we found that *all age groups and both identified genders participated in substantial proportions in multiple types of data citizenship*. Gender and age stratification seemed less relevant for pushing forward the public debate than the influence of various forms of digital experience, such as the contrast between the self-expressive social network involvement and the instrumental online shopping experience. The young (aged 15–24 years) were almost entirely divided between the social netizens and the data citizens – thus disconfirming the definition of the digital natives as dominantly carefree as regards privacy (TNS Opinion & Social 2010). The proportion of young people who were digitally involved and engaged with the GDPR was similar to the proportion of young adults, and higher than the proportions of the other age groups. The more mature generations were, on average, more evenly divided among the clusters, with higher proportions of web citizens and off-line citizens. The latter were typical of people aged 55 years and older (see Table 5).

*Table 5 - Sociodemographic profiles of clusters. Data source: Eurobarometer 91.2*

| | Cluster: | Off-line citizens (row %) | Social netizens (row %) | Web citizens (row %) | Data citizens (row %) | Total (row %) |
|---|---|---|---|---|---|---|
| **Age** | 15 - 24 years | 3.0 | **50.0** | 3.6 | **43.4** | 100.0 |
| | 25 - 39 years | 6.0 | **41.4** | 9.6 | **43.1** | 100.0 |
| | 40 - 54 years | 12.6 | 35.1 | 19.0 | 33.3 | 100.0 |
| | 55 years and older | **44.2** | 18.6 | 24.6 | 12.6 | 100.0 |
| | Total | 22.2 | 32.0 | 17.0 | 28.8 | 100.0 |
| **Gender** | Man | 20.2 | 30.6 | 18.8 | 30.4 | 100.0 |
| | Woman | 24.0 | 33.4 | 15.4 | 27.3 | 100.0 |
| | Total | 22.2 | 32.0 | 17.0 | 28.8 | 100.0 |
| **Age when completed formal education** | No education / Up to 15 years | 57.6 | 19.1 | 15.6 | 7.8 | 100.0 |
| | 16-19 years old | 22.1 | 34.7 | 18.7 | 24.6 | 100.0 |
| | 20 years and older | 9.4 | 30.0 | 19.7 | 40.9 | 100.0 |
| | Still studying | 2.0 | 50.6 | 4.1 | 43.2 | 100.0 |
| | Total | 22.0 | 32.1 | 17.1 | 28.8 | 100.0 |
| **Occupation** | Self-employed | 11.9 | 32.0 | 16.3 | 39.9 | 100.0 |
| | Managers | 3.8 | 25.5 | 16.9 | 53.9 | 100.0 |
| | Other white-collar jobs | 6.8 | 36.5 | 14.7 | 42.1 | 100.0 |
| | Manual workers | 16.0 | 41.0 | 16.5 | 26.5 | 100.0 |
| | House persons | 30.1 | 33.5 | 11.7 | 24.7 | 100.0 |

| Cluster: | Off-line citizens (row %) | Social netizens (row %) | Web citizens (row %) | Data citizens (row %) | Total (row %) |
|---|---|---|---|---|---|
| Unemployed | 18.5 | 46.3 | 12.8 | 22.3 | 100.0 |
| Retired | 49.6 | 16.0 | 25.0 | 9.4 | 100.0 |
| Students | 2.0 | 50.6 | 4.1 | 43.2 | 100.0 |
| Total | 22.2 | 32.0 | 17.0 | 28.8 | 100.0 |

As shown in Figures 1 and 2, higher numbers of data citizens exist in countries such as The Netherlands, Ireland, and the United Kingdom. Over 40% of respondents in these countries are data citizens. In Germany, the Netherlands, and Poland, over 20% of the population is part of the web citizens cluster. Sweden and Cyprus have over 40% of their population in the social netizen type, whereas Bulgaria, Romania, and Greece have the highest rates of off-line citizens, with proportions larger than 33%, due to their lower digitalization in general.

*Fig. 1 - Country distribution for the off-line citizen and social netizens clusters. Data source: Eurobarometer 91.2.*

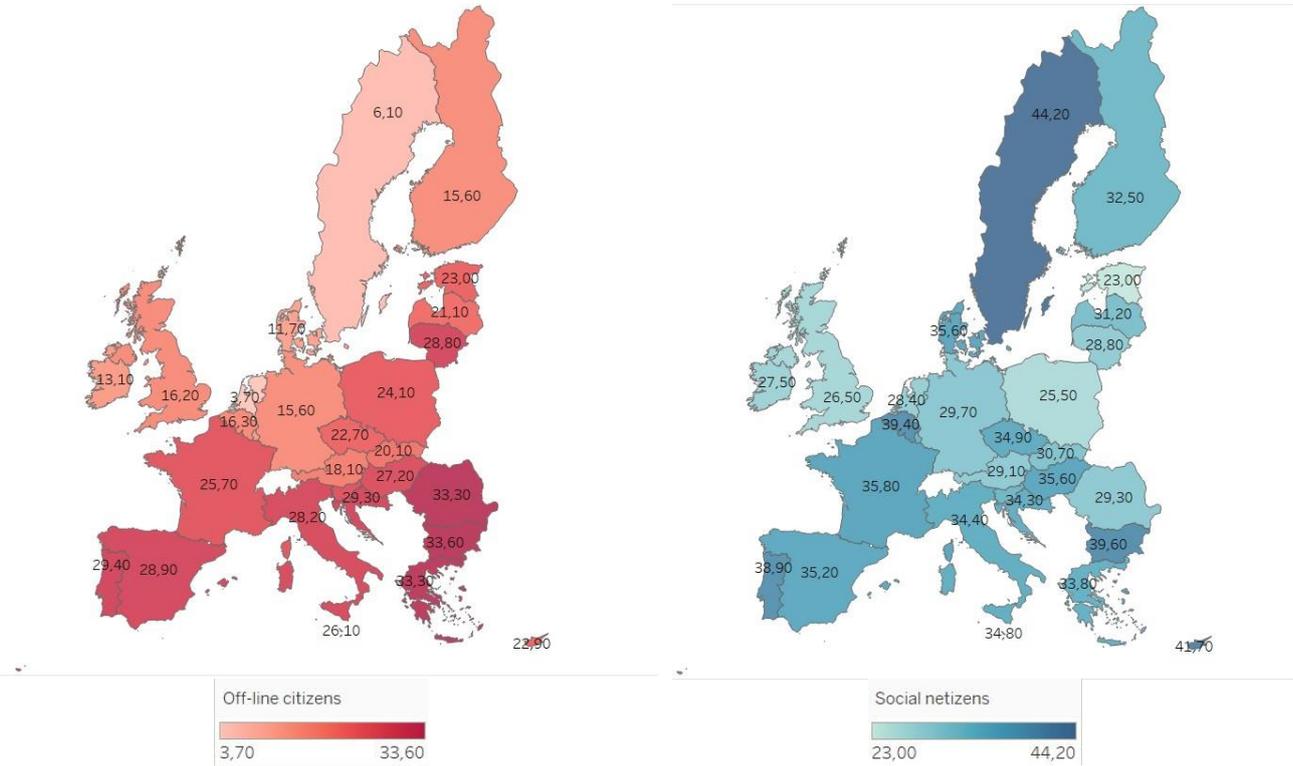

*Fig. 2 - Country distribution for the web citizen and data citizen clusters. Data source: Eurobarometer 91.2.*

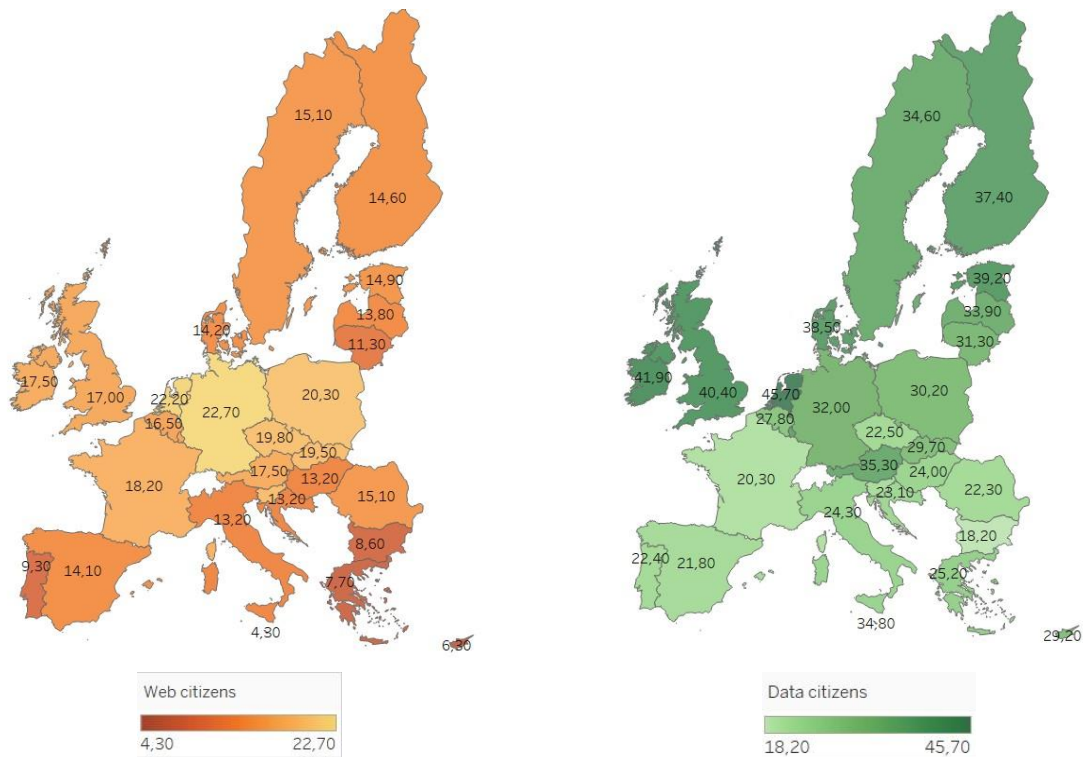

The variability of digital citizenship types at country level (see Figure 1 and Figure 2) was derived from the corresponding national-level variation of digitalization in economy and society. Using an external measure of digitalization, the publicly available EU Digital Economy and Society Index (DESI) computed for all EU27-UK countries (de Montjoye and Houssiau 2020), we observed a strong bivariate correlation between the proportion of "data citizens" within each country and the country index values, especially for the subcomponents of *Internet Use* and *Human Capital.* Digitalization was related to the national-level proportion of data citizens both within the total population and within the younger generations (aged 15–39). At the other end of the continuum, digitalization was strongly and negatively correlated with the country-level proportion of off-line citizens. The proportion of social netizens was also strongly and negatively predicted by economic and social digitalization for entire country-level populations, but not for the subsample of young people, where it seemed that other driving cultural forces were at play. By and large, the growth of GDPR awareness appeared to be in a direct relationship with economic and social digitization, particularly in its influences over the workforce and the daily life (see Table 6).

*Table 6 - Bivariate Pearson correlation coefficients for country-level data citizenship proportions and DESI values. Data source: Eurobarometer 91.2 and the DESI dataset (de Montjoye and Houssiau 2020).*

| | | Connectivity | Human Capital | Use of internet | Integration of Digital Technology | Digital Public Services |
|---|---|---|---|---|---|---|
| **Pearson Correlations for the young population** | **% Off-line citizens** | -0.33 | -0.46 | -0.47 | -0.59 | -0.54 |
| | **% Social netizens** | -0.38 | -0.57 | -0.59 | -0.34 | -0.39 |
| | **% Web citizens** | 0.07 | -0.14 | -0.12 | -0.12 | -0.26 |
| | **% Data citizens** | 0.42 | 0.61 | 0.62 | 0.41 | 0.42 |
| **Pearson correlations** | **% Off-line citizens** | -0.77 | -0.79 | -0.83 | -0.70 | -0.53 |
| | **% Social netizens** | 0.07 | -0.15 | -0.08 | 0.03 | -0.05 |
| | **% Web citizens** | 0.28 | 0.34 | 0.26 | 0.19 | 0.09 |
| for the total population | % Data citizens | 0.60 | 0.73 | 0.77 | 0.60 | 0.55 |

# 4 Conclusions

The GDPR is changing global privacy awareness and practices, but its full potential is still to be fulfilled. People's active involvement in the emerging data-driven society, through data citizenship at all levels of expertise and responsibility, is a condition for this legal apparatus to change the rules of the game. GDPR awareness is stratified by major sociodemographic divisions, of which education, occupation, and generation are the strongest predictors overall, with lower statistical effects for subjective economic well-being, locality size, and gender. Adding to this first layer of basic gradations, digital experience is the strongest predictor of GDPR awareness, with higher levels of digital involvement leading, on average, to higher knowledge of the EU privacy policy. Still, our typological analysis revealed both quantitative and qualitative variations in digital citizenship. Using indicators for digital experience and GDPR awareness, we classified EU27-UK respondents in four clusters, namely, the off-line citizens (22%), the social netizens (32%), the web citizens (17%), and the data citizens (29%). The off-line citizens ranked lowest in internet use and GDPR awareness, the web citizens ranked at about average values, while the data citizens ranked highest in both digital experience and GDPR knowledge and use. The fourth identified cluster, the social netizens, had a discordant profile, with remarkably high social network use, below average online shopping experiences, and low GDPR awareness.

In line with previous research, we found that intense involvement with social networks is a liability for GDPR awareness. The young generations (aged 15–24 and 24–39 years) were relatively evenly divided between the social netizen and the data citizen clusters, with similar proportions of above 40% for each at the EU27-UK level. This indicates that policies should try to remedy the privacy disengagement of the intense social network users, while co-opting the data citizens to ever higher extents into proactive participation in the European data-driven economy and society. When examining *country-level variation in cluster distribution*, we found that *digitalization*, particularly as regards general internet use and human capital, was a strong positive correlate of the proportion of data citizens within national publics, and a negative correlate of the proportion of social netizens and offline citizens. This points to *possible risks of accumulating disadvantages in societies that are less digitized*, compounding lower added value from digital participation with higher risks from personal data exposure and lower compliance with the GDPR. Privacy and data protection notions and skills are yet to be included in school curricula and large-scale public communication across the EU, while *algorithmic awareness*, though closely linked to data-focused debates, is infrequent in the European publics. Our research documents that the general public of European countries is adapting to digitalization and taking note of the new legal instrument of the GDPR, but structural obstacles, such as digital gaps and social media immersion, need to be systematically overcome in order to effectively enforce the rights to private life and the protection of personal data.

**Funding**

This research has been funded by Ministry of Education and Scientific Research, Romania, project PN-III-P4-ID-PCE-2020-1589.

# 5 References

Acquisti A, Adjerid I, Brandimarte L. Gone in 15 seconds: The limits of privacy transparency and control. IEEE Secur Priv. 2013;11(4):72–4.

Acquisti A, Brandimarte L, Loewenstein G. Privacy and human behavior in the age of information. Vol. 347, Science. American Association for the Advancement of Science; 2015. p. 509–14.

Becher SI, Benoliel U. Law in Books and Law in Action: The Readability of Privacy Policies and the GDPR. In: Consumer Law & Economics. Springer; 2020.

Braghin C, Vecchio M Del. Is Pokémon GO Watching You? A Survey on the Privacy-Awareness of Location-Based Apps' Users. In: Proceedings - International Computer Software and Applications Conference. IEEE Computer Society; 2017. p. 164–9.

Brandimarte L, Acquisti A, Loewenstein G. Misplaced Confidences. Soc Psychol Personal Sci. 2013 May;4(3):340–7.

Calvo RA, Deterding S, Ryan RM. Health surveillance during covid-19 pandemic. BMJ. 2020 Apr;369:m1373.

Cheney-Lippold J. We Are Data: Algorithms and The Making of Our Digital Selves. New York: NYU Press; 2017.

Cuevas Á, Cabañas JG, Arrate A, Cuevas R. Does Facebook Use Sensitive Data for Advertising Purposes? Worldwide Analysis and GDPR Impact. 2019 Jul;

Custers B, Dechesne F, Sears AM, Tani T, van der Hof S. A comparison of data protection legislation and policies across the EU. Comput Law Secur Rev. 2018 Apr;34(2):234–43.

Engels B. Digital First, Privacy Second? Digital Natives and Privacy Concerns. In: 17th International Conference e-Society 2019. 2019.

Eubanks V. Automating Inequality: How High-Tech Tools Profile, Police, and Punish the Poor. New York: St. Martin's Press; 2018.

European Commission. Eurobarometer 91.2. Kantar Public, GESIS Data Archive, Cologne. 2019.

European Commission, Kantar Public. Eurobarometer 91.2. ZA7562 Data file Version 1.0.0. Brussels: GESIS Data Archive; 2019.

European Council. Joint European Roadmap towards lifting COVID-19 containment measures. 2020.

Gerber N, Gerber P, Volkamer M. Explaining the privacy paradox: A systematic review of literature investigating privacy attitude and behavior. Vol. 77, Computers and Security. Elsevier Ltd; 2018. p. 226–61.

Goffman E. The Presentation of Self in Everyday Life [Internet]. University Of E, editor. Teacher. Doubleday; 1959. (Anchor books; vol. 21). Available from: http://www.amazon.com/dp/0385094027

Gran AB, Booth P, Bucher T. To be or not to be algorithm aware: a question of a new digital divide? Inf Commun Soc. 2020;

Greenberg A. How Apple and Google Are Enabling Covid-19 Contact-Tracing. WIRED. 2020;

Hallinan D, Friedewald M, McCarthy P. Citizens' perceptions of data protection and privacy in Europe. Comput Law Secur Rev. 2012 Jun;28(3):263–72.

Hu X, Sastry N. Characterising third party cookie usage in the EU after GDPR. In: WebSci 2019 - Proceedings of the 11th ACM Conference on Web Science. New York, New York, USA: Association for Computing Machinery, Inc; 2019. p. 137–41.

Kayes I, Iamnitchi A. Privacy and security in online social networks: A survey. Vols. 3–4, Online Social Networks and Media. Elsevier B.V.; 2017. p. 1–21.

Lerman J. Big Data and Its Exclusions. Stanford Law Rev Online. 2013;66.

Lyon D. Surveillance Capitalism, Surveillance Culture and Data Politics. In: Bigo D, Isin E, Ruppert E, editors. Data Politics: Worlds, Subjects, Rights. Routledge; 2019. p. 64–77.

de Montjoye Y-A, Houssiau F. Can we fight COVID-19 without resorting to mass surveillance? Computational Privacy Group Blog. 2020.

Nissenbaum H. Privacy in Context: Technology, Policy, and the Integrity of Social Life. Stanford: Stanford University Press; 2010.

Nissenbaum H. Contextual integrity up and down the data food chain. Theor Inq Law. 2019;20(1):221–56.

O'Neil C. Weapons of Math Destruction: How Big Data Increases Inequality and Threatens Democracy. Portland, OR: Broadway Books; 2017.

Park S, Humphry J. Exclusion by design: intersections of social, digital and data exclusion. Inf Commun Soc. 2019 Jun;22(7):934–53.

Roberds W, Schreft SL. Data security, privacy, and identity theft: The economics behind the policy debates. Econ Perspect. 2009;33(1).

Rughiniș C, Rughiniș R. Nothing ventured, nothing gained. Profiles of online activity, cyber crime exposure, and security measures of end users in the European Union. Comput Secur. 2014;43:111–25.

Sanchez-Rola I, Dell'Amico M, Kotzias P, Balzarotti D, Bilge L, Vervier PA, et al. Can i opt out yet? GDPR and the global illusion of cookie control. In: AsiaCCS 2019 - Proceedings of the 2019 ACM Asia Conference on Computer and Communications Security. New York, NY, USA: Association for Computing Machinery, Inc; 2019. p. 340–51.

Tene O, Polonetsky J. Privacy in the Age of Big Data: A Time for Big Decisions. Stanford Law Rev Online. 2011;64.

TNS Opinion & Social. Special Eurobarometer 359 "Attitudes on Data Protection and Electronic Identity in the European Union." 2010.

Urban T, Tatang D, Degeling M, Holz T, Pohlmann N. A Study on Subject Data Access in Online Advertising After the GDPR. In: Lecture Notes in Computer Science (including subseries Lecture Notes in Artificial Intelligence and Lecture Notes in Bioinformatics). Springer; 2019. p. 61–79.

Yates S, Carmi E, Pawluczuk A, Lockley E, Wessels B, Gangneux J. Me and My Big Data Report 2020. 2020.

Zuboff S. Big other: Surveillance capitalism and the prospects of an information civilization. J Inf Technol. 2015;

Zuboff S. The age of surveillance capitalism: The fight for a human future at the new frontier of power. Profile Books; 2019.